\documentclass{elsarticle}
\usepackage[usenames,dvipsnames]{color}
\usepackage{hyperref}

\makeatletter
\def\elsartstyle{%
	\def\normalsize{\@setfontsize\normalsize\@xiipt{14.5}}
	\def\small{\@setfontsize\small\@xipt{13.6}}
	\let\footnotesize=\small
	\def\large{\@setfontsize\large\@xivpt{18}}
	\def\Large{\@setfontsize\Large\@xviipt{22}}
	\skip\@mpfootins = 18\p@ \@plus 2\p@
	\normalsize
}
\makeatother

\def\url#1{{\ttfamily\def\/{/\discretionary{}{}{}}#1}}

\begin{document}

\begin{frontmatter}
\title{Perturbation of mass accretion rate, associated acoustic geometry and  stability analysis  }
\author[hri,camk]{Deepika~A~Bollimpalli}
\ead{deepika@camk.edu.pl}
\author[itcp,iucaa]{Sourav~Bhattacharya}
\ead{sbhatta@iucaa.in}
\author[hri]{Tapas~K.~Das}
\ead{tapas@hri.res.in}
\address[hri]{Harish-Chandra research Institute, Chhatnag Road, Jhunsi, Allahabad-211019, India}
\address[camk]{Current affiliation: Nicolaus Copernicus Astronomical Centre, Bartycka 18, 00-716, Warsaw, Poland.}
\address[itcp]{Institute of Theoretical and Computational Physics, Department of Physics\\ University of Crete, 700 13 Heraklion, Greece}
 \address[iucaa]{Current affiliation : Inter-University Centre for Astronomy and Astrophysics (IUCAA) \\Pune University Campus, Pune - 411 007, India.}

\begin{abstract}
\noindent
We investigate the stability of stationary integral solutions of an ideal irrotational fluid in a general static and spherically symmetric background, by studying the profile of the perturbation of the mass accretion rate. We consider low angular momentum axisymmetric accretion flows for three different accretion disk models and consider time dependent and radial linear perturbation of the mass accretion rate. First we show that the propagation of such perturbation can be determined by an effective $2\times2$ matrix, which has qualitatively similar acoustic causal properties as one obtains via the perturbation of the velocity potential.  Next, using this matrix we analytically address the stability issues, for both standing and
travelling wave configurations generated by the perturbation. Finally, based on this general formalism we briefly discuss the explicit example of the Schwarzschild spacetime and compare our results of stability with the existing literature, which instead address this problem via the perturbation of the velocity potential.
\end{abstract}

\end{frontmatter}


\section{Introduction}
\noindent
To obtain reliable spectral signatures of astrophysical black holes using a set of 
stationary transonic accretion solutions, it is necessary to ensure that such 
integral solutions are stable under perturbation~\cite{kato-book}, at least for an 
astrophysically relevant time scale. In~\cite{paper-I}, such stability was argued via demonstrating the natural emergence of the acoustic analogue geometry through perturbation 
of the mass accretion rate for radial Bondi flows with spherical symmetry~\cite{michel, Bondi1952}. Here we wish to extend this perturbation scheme to accommodate low angular momentum axisymmetric flows, assuming the fluid to be inviscid, irrotational and non self-gravitating, with three different disk models the axisymmetric accretion flow can have. Some preliminary results in this direction can be seen in~\cite{Ananda:2014jza}.

On the analytical front, stationary 
flow solutions for the low angular momentum inviscid accretion 
has extensively been studied in the literature, see, e.g.~\cite{lt80}-\cite{das-czerny-2012-new-astronomy} and references therein.
Numerical works have also been reported for such flow configurations in~\cite{spon-molt}-\cite{okuda2} and references therein.

In this context it should be emphasized that 
the concept of low angular
momentum advective flow (where the inviscid assumption is justified)
is not a theoretical abstraction.
Sub-Keplerian flows are observed in nature.
Such flow configurations may be observed
for detached binary systems
fed by accretion from OB stellar winds~\cite{ila-shu,liang-nolan},
semi-detached low-mass non-magnetic binaries~\cite{bisikalo},
and supermassive black holes fed
by accretion from slowly rotating central stellar clusters~\cite{ila,ho} (see also the references therein).
Even for a standard Keplerian
accretion disc, turbulence may produce such low angular momentum flow, see, e.g.~\cite{igu} and references therein. Moreover, given a background spacetime, it is natural to expect a critical value of the specific angular momentum of the flow, below which there would not be any Keplerian orbits. 

In addition to the analysis of the stationary transonic solutions, 
the stability properties of such flow has also been performed in various 
works,
see, e.g.~\cite{M1980}-{\cite{bose-sengupta-ray} and references therein.

In the present work, we study the linear perturbation of the stationary 
transonic black hole accretion solutions through its connection to the emergence of the sonic 
geometry embedded within spacetime 
characterizing the background stationary flow (see~\cite{M1980, unr81, bar05}, and references therein,
for detailed discussion about analogue gravity phenomena). 

The emergent analogue acoustic geometry in accretion astrophysics via perturbation of the potential of irrotational velocity flow and related stability issues was first studied extensively in~\cite{M1980}, for spherical accretion.
To the best of our knowledge, as of now the acoustic geometry associated with usual analogue gravity models has been obtained by perturbing the corresponding velocity potential of the background fluid flow. Instead, in the present work we discuss the sonic causal structure 
by perturbing the mass accretion rate associated with the infalling matter.
The motivation behind this is obvious -- the mass accretion rate is an astrophysically 
relevant and measurable quantity, and hence it is interesting to determine its profile perturbatively. Moreover, mass accretion rate is associated with both the density field and velocity field, which provide the full description of the flow. Therefore, perturbing the mass accretion rate leads to a wave equation that can shed light on stability of both the fields.
In~\cite{paper-I}, such connection was demonstrated for spherical accretion, generalizing the non-relativistic results of~\cite{jkb-arnab-tapan-nabajit}. In this work, we wish to extend these earlier results for axisymmetric matter flow with nontrivial disk structures in general static and spherically symmetric spacetimes.


However, there is a crucial difference between the acoustic geometry we derive here with the same 
obtained via the perturbation of the velocity potential. 
We shall ignore
any non-axisymmetric features of the perturbation corresponding to the mass accretion rate, 
and assume that the accretion rate is a function of the radial and the time coordinates 
only. 
Accordingly, the internal geometry through which the perturbation propagates has dimension two, 
instead of three or four. Nevertheless, we will see that the qualitative features regarding the causal structures 
of these two acoustic geometries remain the same, even though the mass accretion rate and the velocity potential are two very distinct quantities. 
To the best of our knowledge, such connection 
has not been reported in the existing literature. 

For axisymmetric accretion, infalling matter can have three different geometric configurations -- 
the conical, the constant height and the vertical equilibrium (at least for thin acrretion disks) models 
(see, e.g.~\cite{swagata} and references therein). 
In subsequent sections we shall derive the general relativistic acoustic geometry through which the 
linear perturbation (corresponding to the mass accretion rate) propagates, for all these three models. 
We then briefly address the stability properties of the axisymmetric matter flows and reestablish the chief qualitative features reported earlier using different methods~\cite{M1980}, for spherical accretion. 


We shall use mostly positive signature for the metric and will set $c=1=G$ hereafter. We outlined the paper in the following manner. In section 2 we briefly discuss the preliminaries and the equations describing the relativistic axisymmetric fluid flow in static and spherically symmetric spacetime for all the three disk models considered. In section 3, we first discuss the stationary solutions of the flow and derive the critical point condition. Following this we perform the linear mass accretion rate perturbation and derive the effective $2\times 2$ matrix through which the perturbation propagates. Using this in section 4, we address the stability issues by deriving the profile of the perturbation of the mass accretion rate. In section 5, we summarize our work with an outlook. 

\section{The basic constructions}
\noindent
We shall briefly mention here the basic ingredients and assumptions necessary for our calculations.
Let us start with the metric for a general static and spherically symmetric spacetime
\begin{eqnarray}
ds^{2}= -g_{tt}(r) dt^{2}+g_{rr}(r)dr^{2}+ r^2d\Omega^{2},
\label{e1}
\end{eqnarray}
where $d\Omega^2$ is the line element of a unit 2-sphere.

Let us take an ideal and inviscid fluid with the energy-momentum tensor,
\begin{eqnarray}
T_{\mu \nu} = \left( \epsilon + p \right) v_{\mu} v_{\nu} + pg_{\mu \nu},
\label{e2}
\end{eqnarray} 
where $\epsilon$ and $p$ are respectively the mass-energy density and pressure, $v^{\mu}$ is the fluid's four velocity normalized as $v^{\mu}v_{\mu}=-1$.

We shall consider axisymmetric radial flow on the equatorial plane, hence $v^{\theta}=0$. The normalization condition thus provides
\begin{eqnarray}
v^{t}= \sqrt{\frac{1+g_{rr} v^{2}+g_{\phi \phi} (v^{\phi})^2}{g_{tt}}},
\label{e3}
\end{eqnarray}
where we have written $v^{r} \equiv v$.
We assume that the fluid obeys the adiabatic equation of state $p = k \rho^{\gamma}$, where $\gamma$ is adiabatic index and $\rho$ is the fluid density. The specific enthalpy $h$ of the fluid is given by $h = \frac{\epsilon+p}{\rho}$, so that
\begin{eqnarray}
dh = T d \left( \frac{S}{\rho} \right) + \frac{dp}{\rho},
\label{e4}
\end{eqnarray}
where $T$ and $S$ are the temperature and entropy of the fluid respectively. Under isoentropic conditions, we can define the speed of the sound, $c_{\rm s}$ to be
\begin{eqnarray}
c_{\rm s}^{2} = \frac{\partial p}{\partial \epsilon}\Bigg\vert_{dS=0}.
\label{e5}
\end{eqnarray}  
Since the flow is axisymmetric, the equation of continuity, $\nabla_{\mu} \left(\rho v^{\mu} \right)=0$ provides
\begin{eqnarray}
\partial_{t}\left(\rho v^{t} \sqrt{-{g}}H_{\theta} \right)+\partial_{r}\left(\rho v \sqrt{-{g}}H_{\theta} \right)=0,
\label{e6}
\end{eqnarray}
where we have used $v^{\theta}=0$ and 
$H_{\theta}$ is a function of the local flow thickness, which depends upon the model of the accretion we are choosing (see e.g.~\cite{swagata} and references therein for details).
In other words, $H_{\theta}$ is an appropriate weight function required
to define the flux of mass falling onto the accretor,
implicitly providing the detailed structure of the accretion along a direction normal to the 
equatorial plane ($\theta=\pi/2$). Thus the weight function appears as the consequence of 
the averaging over $\theta=\pi/2$, and hence use of such function 
effectively permits us to construct the differential equations governing the flow profile 
on the equatorial plane by considering the stratification (through averaging) due to the 
non-equatorial distribution of any thermodynamic flow variable. 

For example, if the accretion flow makes a cone or a sphere (with the accretor at the centre), which subtends a constant solid angle at the origin for all radial values, $H_{\theta}$ will be a constant. For radial flow from all directions, this factor will clearly be unity, and less than unity otherwise. Likewise, for flow with constant thickness, we may have $H_{\theta}=\frac{H}{r}$, where $H$ is a constant. This ensures that the solid angle subtended by the edges of the flow lines will decrease with radial distance. The most nontrivial 
flow geometry is the flow in hydrostatic equilibrium along the vertical direction, for which $H_{\theta}$ is a function of $r$ as well as of certain flow variables, except the radial velocity. It is also assumed that $H_{\theta}$ is not an explicit function of time. It is thus obvious that flow with constant height and the conical flow can be obtained from vertical equilibrium configuration imposing certain limiting conditions. We shall consider all these three disk models unitedly in our analysis below.

 We also note that the weight function can also effectively be understood by replacing $d\Omega^2$ in the metric by $H_{\theta}d\Omega^2$, so that when one integrates the continuity equation, the factor $H_{\theta}$ in the integrand takes care of the geometric configuration of the off-equatorial infalling matter.

The conservation equation for the energy-momentum tensor, $\nabla_{\mu} T^{\mu \nu} = 0$ along with the continuity equation and using certain thermodynamic relations, provides
\begin{eqnarray}
v^{\nu} \partial_{\nu} v^{\mu} + v^{\nu} v^{\lambda} \Gamma^{\mu}_{\nu \lambda} + \frac{c_{\rm s}^2}{\rho}\left( v^{\mu} v^{\nu} + g^{\mu \nu} \right) \partial_{\nu} \rho = 0.
\label{e8}
\end{eqnarray}
For $\mu=t,r, \phi$, we obtain the conservation equations for the energy, the radial and the orbital angular momenta respectively.
\begin{eqnarray}
v^{t} \partial_{t} v^{t} + \frac{c_{\rm s}^2}{\rho}\left(  (v^t)^2- g^{tt} \right)\partial_{t}\rho+ v \partial_{r} v^{t} + g^{tt}\left(\partial_{r}g_{tt}\right)v v^{t} + \frac{c_{\rm s}^2}{\rho} vv^t \partial_{r}\rho = 0,
\label{e9}\\
v^{t} \partial_{t} v + \frac{c_{\rm s}^2}{\rho} vv^{t} \partial_{t} \rho + v \partial_{r} v + \frac{1}{2g_{rr}} \left[ \left(\partial_{r}g_{tt} \right)(v^t)^2+\left(\partial_{r}g_{rr} \right)v^{2}-\left(\partial_{r}g_{\phi \phi} \right)(v^{\phi})^2)\right] + \frac{c_{\rm s}^2}{\rho} \left[ v^{2} +g^{rr}\right] \partial_{r}\rho= 0, 
\label{e10}\\
v^{t} \partial_{t} v^{\phi} + \frac{c_{\rm s}^2}{\rho}v^{\phi} v^{t}\partial_{t}\rho+ v \partial_{r} v^{\phi} + g^{\phi \phi}\left(\partial_{r}g_{\phi \phi}\right)v v^{\phi} + \frac{c_{\rm s}^2}{\rho} vv^{\phi} \partial_{r}\rho = 0,
\label{e11}
\end{eqnarray}
where here and hereafter $g_{\phi\phi}$ will always stand for $g_{\phi\phi}(\theta=\pi/2)$.

We assume that the flow is irrotational, so that
\begin{eqnarray}
\partial_{\mu} \left( h v_{\nu} \right) - \partial_{\nu} \left( h v_{\mu} \right) = 0.
\label{e12}
\end{eqnarray}
Choosing different free indices, we obtain 
\begin{eqnarray}
\partial_{t} v_{r} - \partial_{r} v_{t} &=& \frac{c_{\rm s}^{2}}{\rho}\left[ v_{t} \partial_{r} \rho - v_{r} \partial_{t}  \rho  \right],\nonumber\\
\label{e13}
\partial_{t}v^{\phi} &=& -\frac{v^{\phi} c_{\rm s}^{2}}{\rho} \partial_{t}\rho,\nonumber\\
\label{e14}
\partial_{r}v^{\phi} &=& -\frac{v^{\phi} c_{\rm s}^{2}}{\rho} \partial_{r}\rho - \frac{v^{\phi}\partial_{r}g_{\phi \phi}}{g_{\phi \phi}}.
\label{e15}
\end{eqnarray}
Using the normalization and the irrotational conditions, Eq.~(\ref{e10}) can be re-phrased as
\begin{eqnarray}
v^{t}\partial_{t}v +\frac{c_{\rm s}^{2}}{\rho} v v^{t} \partial_{t} \rho + \frac{c_{\rm s}^{2}}{\rho} \left(\frac{g_{tt}(v^t)^2}{g_{rr}} \right)\partial_{r} \rho+ \partial_{r}\left(\frac{g_{tt}(v^t)^2 }{2 g_{rr}}\right) + \left(\frac{g_{tt} (v^t)^2}{2g_{rr}}\right)\frac{\partial_{r}\left(g_{tt}g_{rr}\right)}{g_{tt}g_{rr}}=0.
\label{e16}
\end{eqnarray}
$\partial_t \equiv 0$ corresponds to the stationary state. When we perturb the radial equation, we shall include time dependence, but will ignore `$\phi$' dependence throughout.

We next derive an expression for the derivative of the weight function $H_{\theta}$. Note that $c_{\rm s}^2$, $\rho$ and $p$ are interrelated by more than one thermodynamic relations. So, without any loss of generality, we may assume $H_{\theta}\equiv H_{\theta}(r, v^{\phi}, \rho)$. The general form of the irrotational condition for $v^{\phi}$ is (\ref{e15})
\begin{eqnarray}
\partial_{x}v^{\phi} &=& -\frac{v^{\phi} c_{\rm s}^{2}}{\rho} \partial_{x}\rho - \frac{v^{\phi}\partial_{x}g_{\phi \phi}}{g_{\phi \phi}}.
\label{gen}
\end{eqnarray}
where, $x$ is either $t$ or $r$. Using the chain rule for the partial differentiation and Eq.~(\ref{gen}), we have 
\begin{eqnarray}
 \frac{1}{H_{\theta}}\frac{d H_{\theta}}{dx} = \alpha+ \beta \frac{\partial_{x}\rho}{\rho},
\label{e22}
\end{eqnarray}
where $\displaystyle \alpha=\frac{\partial \ln H_{\theta}}{\partial x}-v^{\phi}\left(\frac{\partial \ln H_{\theta}}{\partial v^{\phi}}\right)\frac{\partial\ln g_{\phi\phi}}{\partial x}$ and $\displaystyle \beta=\rho \frac{\partial \ln H_{\theta}}{ \partial \rho}- v^{\phi} c^2_{\rm s}\left(\frac{\partial \ln H_{\theta}}{ \partial v^{\phi}}\right) $. Since it has already been assumed that $H_{\theta}$ has no explicit time dependence, we always have $\partial_t H_{\theta}=0$.

We recall now the general expression for $H_{\theta}$ satisfying the vertical equilibrium condition~\cite{Abramowicz:1996ap}, 
\begin{eqnarray}
 -\frac{p}{\rho}+\zeta(r) H^2_{\theta}v^2_{\phi}=0,
\label{e22'}
\end{eqnarray}
where $\zeta(r)$ is independent of fluid variables. We substitute from here the expression for $H_{\theta}$
into the expression for $\beta$ appearing below Eq.~(\ref{e22}), to get 
\begin{equation}
\beta = \frac{\gamma- 1}{2}+c_s^2,
\label{e22''}
\end{equation}
which is always positive, since $\frac43\leq\gamma\leq \frac53$~\cite{weinberg}. Hereafter, we shall always consider $\beta\geq 0$, where the equality holds for the conical and the constant height models, since none of them depends upon the flow variables.

\section{Propagation of the linear perturbation}
\label{5.22}

\subsection{The stationary solutions}
\label{5.21}
\noindent
We shall first discuss the stationary solutions of the flow equations constructed in the previous section.
Integration of the spatial part of the continuity equation, Eq. (\ref{e6}), provides
\begin{eqnarray}
\rho_{0} v_{0} \sqrt{-\widetilde{g}} H_{\theta_{0}}  = {\rm constant},
\label{e17}
\end{eqnarray}
where $\widetilde{g}=- r^4g_{tt}g_{rr}$ and the subscript `0' denotes stationary value. We multiply this by the element of solid angle $d\Omega=\sin^2\theta d\theta d\phi$, and integrate to define the mass accretion rate,
\begin{eqnarray}
\Omega \rho_{0} v_{0} \sqrt{-\widetilde{g}} H_{\theta_{0}}  = {\rm constant}=\Psi_0~(\rm say),
\label{e17'}
\end{eqnarray}
It is clear that we can absorb the factor $\Omega$ without loss of any generality. Hence we shall call $\Psi_0= \rho_{0} v_{0} \sqrt{-\widetilde{g}} H_{\theta_{0}}$ to be the mass accretion rate.

Setting the time derivatives in the energy conservation equation, Eq.~(\ref{e9}) to zero, we find
\begin{eqnarray}
h v_{0t} = -\cal{E},
\label{e18} 
\end{eqnarray}
where ${\cal E}$ is a constant along the flow line and is identified as the specific energy. We also note, using Eq.s~(\ref{e15}) and (\ref{e18}), that the specific angular momentum, the parameter $\displaystyle \lambda = -\frac{v_{0\phi}}{v_{0t}}=\frac{ l}{{\cal E}}$ is a constant along the flow line as well. 
 
Similarly, setting the time derivatives in the radial equation, Eq.~(\ref{e16}), to zero, provides
\begin{eqnarray}
-\frac{2c_{s_{0}}^{2}}{\rho_{0}} \partial_{r}\rho_{0}= \left( \frac{g_{rr}}{g_{tt} (v_{0}^{t})^{2}} \right) \partial_{r} \left( \frac{g_{tt} (v_{0}^{t})^{2}}{g_{rr}} \right) +  \frac{\partial_{r} ( g_{rr} g_{tt})}{ g_{rr} g_{tt}}.
\label{e19}
\end{eqnarray}
We now transform to a local frame following~\cite{Gammie:1997ct},
\begin{eqnarray}
v_0^{t} &=& \sqrt{\frac{g_{\phi \phi}}{g_{tt}\left(g_{\phi \phi} - g_{tt} \lambda^{2} \right)}}\sqrt{\frac{1}{1-u_0^{2}}},\nonumber \\
v_0 &=& u_0\sqrt{\frac{1}{g_{rr}\left(1-u_0^{2}\right)}},\nonumber\\
v_0^{\phi} &=& \lambda \sqrt{\frac{g_{tt}}{ g_{\phi\phi
}\left(g_{\phi \phi} - g_{tt} \lambda^{2} \right)}} \sqrt{\frac{1}{1-u_0^{2}}}.
\label{e20}
\end{eqnarray}
Substituting Eq.s~(\ref{e20}) into Eq.s~(\ref{e18}) and (\ref{e19}), we rewrite the stationary solutions in terms of $u_0$. Then considering the radial derivative of Eq.~(\ref{e18}) and using Eq.s~(\ref{e19}), (\ref{e22}), we find
\begin{eqnarray}
\frac{du_{0}}{dr} = \frac{u_{0}\left(1-u_{0}^{2}\right)\left[\left\lbrace \partial_{r} \ln \left( \frac{-\widetilde{g} }{g_{rr}} \right) +2\alpha \right \rbrace c_{\rm s_{0}}^{2} -(1+\beta)\partial_{r} \ln \left( \frac{g_{tt}g_{\phi \phi}}{g_{\phi \phi}-\lambda^{2}g_{tt}} \right)\right]}{2\left[u_{0}^{2}\left(1+\beta\right)-c_{\rm s_{0}}^{2} \right]}.
\label{e21}
\end{eqnarray}
We note that the above equation has critical points, where both the denominator and the numerator vanish simultaneously in order for the flow to be transonic. This gives us the condition  
\begin{eqnarray}
u_{0}^{2} \vert_{c}=\frac{c_{\rm s_{0}}^{2}\vert_{c}}{(1+\beta_c)} = \frac{\partial_{r} \ln \left( \frac{g_{tt}g_{\phi \phi}}{g_{\phi \phi}-\lambda^{2}g_{tt}} \right)}{\partial_{r} \ln \left( \frac{-\widetilde{g}}{g_{rr}} \right) +2\alpha_c},
\label{e23}
\end{eqnarray}
where the subscript `$c$' stands for the critical point. For the flow geometries in which the thickness of the flow depends on the flow variables, $\beta>0$ and the critical point is different from the sonic point. In particular, the denominator of Eq.~(\ref{e23}) shows that the sonic point ($u_0^2=c_{\rm s_0}^2$) is reached when $u_{0}^{2}\left(1+\beta\right)-c_{\rm s_{0}}^{2}=\beta u_0^2$. Since outside the critical point, $u_{0}^{2}\left(1+\beta\right)-c_{\rm s_{0}}^{2}<0$, it is clear that the sonic point is always located at smaller value of the radial coordinate than the critical point. 

Once the position of the critical point is known, the corresponding sonic point can be obtained by numerically integrating the differential equation describing the space gradient of the radial velocity $\frac{du}{dr}$, see e.g.~\cite{Pu:2012rv} and references therein for further details.


When we substitute for the metric explicitly into Eq.s~(\ref{e21}), (\ref{e23}), we recover the result for the Schwarzschild spacetime~\cite{Tarafdar:2013oqa}.
 
For a constant height model, $\beta=0,~\alpha = -\frac{1}{r}$ (Eq.~(\ref{e22})) and for the conical model, since $H_{\theta}$ is constant, we have $\beta=0=\alpha$. Also, setting $\lambda=0=\beta$, we recover the results for the spherical Bondi flow~\cite{Bondi1952}. Using these ingredients, in the following we shall derive the time dependent perturbation equations and demonstrate the natural emergence of  of the acoustic causal structure.


\subsection{The acoustic causal structure via this alternative approach}
\noindent
Let us begin by considering the  linear perturbation scheme
\begin{eqnarray}
v(r,t) &=& v_{0}(r)+v'(r,t),
\label{a1}\\
v^{\phi} (r,t) &=& v_{0}^{\phi} (r) + v^{\phi'} (r,t),\label{a2}\\
\rho(r,t) &=& \rho_{0}(r)+\rho'(r,t), 
\label{e25}\\
v^{t'}(r,t) &=& \frac{g_{rr}v_{0}v'+g_{\phi \phi}v_{0}^{\phi}v^{\phi'}}{g_{tt}v_{0}^{t}},
\label{e26}
\end{eqnarray}
where the subscript `$0$' stands for the stationary state and in the last equation we have used the normalization of the velocity.

Let us define a variable $\Psi = \rho v \sqrt{-\widetilde{g}} H_{\theta}$ which, as one can see from Eq.~(\ref{e17'}), becomes identical with the mass accretion rate at its stationary value (apart from a geometric constant $\Omega$). 
Linear expansion of $\Psi(r, t)$ provides
\begin{eqnarray}
\Psi(r,t) =\Psi_0(r)+\Psi'(r,t)= \rho_0v_{0} H_{\theta_{0}} \sqrt{-\widetilde{g}}+\left( \rho'v_{0} H_{\theta_{0}}+\rho_{0}v' H_{\theta_{0}} + \rho_{0} v_{0} H_{\theta}' \right) \sqrt{-\widetilde{g}}.
\label{e27}
\end{eqnarray}
$H_{\theta}$ can be a function of the flow variables, except for the radial velocity~\cite{swagata} and hence the perturbation of certain flow variables induces time dependence on $H_{\theta}$. From Eq.~(\ref{e22}), we can write
\begin{eqnarray}
\frac{d \ln H_{\theta}'}{dt} = \beta \frac{\partial_{t} \rho'}{\rho_0},
\label{e28}
\end{eqnarray}
where $\beta$ is a function of all the time independent terms.
Substituting for the perturbed quantities Eq.s~(\ref{a1})-(\ref{e28}) into the continuity equation (\ref{e6}), we get
\begin{eqnarray}
-\frac{\partial_{r} \Psi'}{\Psi_{0}} =  \frac{1}{g_{tt}v_{0}^{t}v_{0}}\left[ \left(g_{rr}v_{0}^{2}\right)\frac{\partial_{t} v'}{v_{0}} + \left(g_{tt}(v_{0}^t)^2(1+\beta)-g_{\phi \phi}(v_{0}^{\phi})^{2}c_{\rm s_{0}}^{2}\right) \frac{\partial_{t} \rho'}{\rho_{0}} \right].
\label{e29}
\end{eqnarray}
Also, taking the time derivative of Eq.~(\ref{e27}) and using Eq.~(\ref{e28}), we obtain
\begin{eqnarray}
\frac{\partial_{t} \Psi'}{\Psi_{0}} = (1+\beta) \frac{ \partial_{t} \rho'}{\rho_{0}} +  \frac{\partial_{t} v'}{v_{0}}. 
\label{e30}
\end{eqnarray}
Solving Eq.s~(\ref{e29}) and (\ref{e30}), we can express the derivatives of $\rho'$ and $v'$ solely in terms of $\Psi'$,
\begin{eqnarray}
\frac{\partial_{t} v'}{v_{0}} &=& \frac{1}{\Lambda}\left[ \frac{\partial_{t} \Psi'}{\Psi_{0}} + g_{tt}v_{0}v_{0}^{t}(1+\beta) \frac{\partial_{r} \Psi'}{\Psi_{0}} \right],
\label{e31} \\
\frac{\partial_{t}\rho '}{\rho_{0}} &=& -\frac{1}{\Lambda}\left[ g_{rr}v_{0}^{2}\frac{\partial_{t} \Psi'}{\Psi_{0}} + g_{tt}v_{0}v_{0}^{t} \frac{\partial_{r} \Psi'}{\Psi_{0}} \right],
\label{e32}
\end{eqnarray}
where
\begin{eqnarray}
\Lambda = (1+\beta)+(1+\beta-c_{\rm s_{0}}^{2})g_{\phi \phi}(v_{0}^{\phi})^{2}. 
\label{e32'}
\end{eqnarray}
We now substitute for the perturbed quantities into the radial equation, Eq.~(\ref{e16}), to get
\begin{eqnarray}
\left(\frac{g_{rr}v_{0}}{g_{tt}v_{0}^{t}}\right) \frac{\partial_{t}v'}{v_{0}}+\left(\frac{g_{rr}v_{0}c_{\rm s_{0}}^{2}}{g_{tt}v_{0}^{t}}\right)\frac{\partial_{t}\rho'}{\rho_{0}}+\partial_{r}\left( \frac{c_{\rm s_{0}}^{2}\rho'}{\rho_{0}}+\frac{v^{t'}}{v_{0}^{t}}\right) = 0.
\label{e33}
\end{eqnarray}
Taking the time derivative of the above equation and substituting it for the time derivatives of $\rho'$ and $v'$ from Eq.s~(\ref{e31}) and (\ref{e32}), we obtain
\begin{eqnarray}
\partial_{t}\left[ \frac{g_{rr}v_{0}}{v_{0}^{t}\Lambda} \left( \frac{ c_{\rm s_{0}}^{2} +\lbrace (1+\beta)- c_{\rm s_{0}}^{2} \rbrace g_{tt}(v_{0}^{t})^{2}}{g_{tt}} \right) \partial_{t} \Psi' \right]
+ \partial_{t}\left[  \frac{g_{rr}v_{0}}{ v_{0}^{t}\Lambda}  \left(v_{0}v_{0}^{t}\lbrace (1+\beta)- c_{\rm s_{0}}^{2} \rbrace\right) \partial_{r} \Psi'\right] + \nonumber \\
 \partial_{r}\left[ \frac{g_{rr}v_{0}}{ v_{0}^{t}\Lambda}  \left(v_{0}v_{0}^{t}\lbrace (1+\beta)- c_{\rm s_{0}}^{2} \rbrace\right) \partial_{t} \Psi'\right]+ \partial_{r}\left[  \frac{g_{rr}v_{0}}{v_{0}^{t}\Lambda} \left( \frac{ -c_{\rm s_{0}}^{2} +\lbrace (1+\beta)- c_{\rm s_{0}}^{2} \rbrace g_{rr}v_{0}^{2}}{g_{rr}} \right) \partial_{r} \Psi'\right]=0,
\label{e34}
\end{eqnarray}
so that we may readily identify a symmetric matrix $f^{\mu\nu}$ defining the 
corresponding spacetime structure through which the perturbation of the mass accretion rate $\Psi'$ propagates,
\begin{eqnarray}
f^{\mu \nu} \equiv 
\frac{g_{rr}v_{0}c_{\rm s_{0}}^{2}}{v_{0}^{t}\Lambda} \left[ \begin{array}{cc}
 -g^{tt} +\left( 1- \frac{1+\beta}{c_{\rm s_{0}}^{2}} \right) (v_{0}^{t})^{2} & v_{0}v_{0}^{t}\left( 1- \frac{1+\beta}{c_{\rm s_{0}}^{2}} \right) \\ 
\\ v_{0}v_{0}^{t}\left( 1- \frac{1+\beta}{c_{\rm s_{0}}^{2}} \right) &  g^{rr} +\left( 1-\frac{1+\beta}{ c_{\rm s_{0}}^{2}} \right) v_{0}^{2} \\
 \end{array} \right], 
\label{metric1}
\end{eqnarray}
and the equation for the perturbed mass accretion rate takes a compact form,
\begin{eqnarray}
\partial_{ \mu} \left( f^{\mu \nu} \partial_{ \nu} \Psi' \right) = 0.
\label{e35}
\end{eqnarray}
Substituting Eq.s~(\ref{e20}) into the expression for $f^{\mu\nu}$ gives
\begin{eqnarray}
f^{\mu \nu} \equiv 
 \varpi \left[ \begin{array}{cc}
\sqrt{\frac{g_{rr}}{g_{\phi \phi}g_{tt}}}\left(\frac{g_{\phi \phi}\left(u_0^2-\frac{1+\beta}{c_{s_0}^2} \right)+\lambda^2 g_{tt}(1-u_0^2)}{\sqrt{g_{\phi \phi}-\lambda^2g_{tt}}} \right) & u_0\left( 1- \frac{1+\beta}{c_{\rm s_{0}}^{2}} \right) \\ 
\\ u_0\left( 1- \frac{1+\beta}{c_{\rm s_{0}}^{2}} \right) & \sqrt{\frac{g_{tt}(g_{\phi \phi}-\lambda^2 g_{tt})}{g_{rr}g_{\phi \phi}}}\left(1 -\frac{u_0^2(1+\beta)}{c_{\rm s_0}^2} \right)  \\
 \end{array}\right]
\label{metric2}
\end{eqnarray}
where $ \varpi = \frac{u_0c_{s_0}^2(g_{\phi \phi}-\lambda^2 g_{tt})}{(1+\beta)[g_{\phi \phi}(1-u_0^2)+\lambda^2 g_{tt} u_0^2]-\lambda^2 g_{tt}c_{s_0}^2}$. The inverse $f_{\mu\nu}$, of $f^{\mu\nu}$ is given by
\begin{eqnarray}
f_{\mu \nu} \equiv {\frac{1}{u_0 (1-u_0^2)}}\sqrt{\frac{g_{\phi \phi}}{g_{\phi \phi}-\lambda^2 g_{tt}}} \left[ 
\begin{array}{cc}
 \sqrt{\frac{g_{tt}}{g_{rr}}}\left(\frac{u_0^2(1+\beta)}{c_{s_0}^2}-1 \right)& u_0\sqrt{\frac{g_{\phi \phi}}{g_{\phi \phi}-\lambda^2g_{tt}}}\left( 1- \frac{1+\beta}{c_{\rm s_{0}}^{2}} \right) \\ 
 u_0\sqrt{\frac{g_{\phi \phi}}{g_{\phi \phi}-\lambda^2g_{tt}}}\left( 1- \frac{1+\beta}{c_{\rm s_{0}}^{2}} \right) & \sqrt{\frac{g_{rr}}{g_{tt}}}\left(\frac{g_{\phi \phi}\left(\frac{1+\beta}{c_{s_0}^2}-u_0^2 \right)-\lambda^2 g_{tt}(1-u_0^2)}{g_{\phi \phi}-\lambda^2g_{tt}} \right)
\end{array}\right],
\label{metric3}
\end{eqnarray}
we once again recall that $g_{\phi\phi}$ is evaluated on $\theta=\pi/2$.

In analogy to the Kerr spacetime, we may define the acoustic ergo region as a region in spacetime where the stationary Killing vector $\xi$ becomes spacelike ~\citep{B1999}. The boundary of this ergo region is defined as a hypersurface called the stationary limit surface on which the magnitude of this killing vector with respect to the acoustic metric vanishes. This condition is equivalent to 
\begin{eqnarray}
 f_{tt} = 0 \Rightarrow u_{0}^2 = \frac{c_{s_0}^2}{(1+\beta)}
\label{ec}
\end{eqnarray}
Recollect from Eq.~(\ref{e23}), that the above equation is nothing but the condition for critical point. This indicates that the propagation of $\Psi'$ has qualitatively similar acoustic causal structure as the velocity potential~\cite{M1980, unr81}.\footnote{To the best of our knowledge, the existing literature on the velocity potential approach always considers $\beta=0$.} 

To see this more clearly, let us compare our result with the existing literature in a bit more details. Firstly, the part of $f^{\mu\nu}$ in Eq.~(\ref{metric1}) written within the matrix manifestly behaves as a rank 2 tensor, defined on the `$t-r$' plane. The overall conformal factor $\left(=\frac{g_{rr}v_{0}c_{\rm s_{0}}^{2}}{v_{0}^{t}\Lambda}\right)$ is not a tensor, due to appearances of the multiplied velocity and metric components. Accordingly, the symmetric matrix $f^{\mu\nu}$ is not a tensor as a whole, unlike the one gets for the perturbation of the velocity potential~\cite{M1980, unr81}. This mismatch should be attributed to the fact that the mass accretion rate, being dependent upon the component of the velocity and determinant of the metric (Eq.s~(\ref{e17}), (\ref{e17'})), is manifestly not a scalar quantity, unlike the velocity potential.

The 2-tensorial part of $f^{\mu\nu}$ is similar to that one gets via the perturbation of the velocity potential~\cite{M1980, unr81, bar05, Barcelo:2004wz}, for conical flow of the fluid, $\beta=0$. The chief difference between our result and the velocity potential approach is the fact that our geometry is two dimensional, since we have taken $\Psi'$ to be independent of the azimuthal angle $\phi$, while information about the axisymmetric flow being contained in $\lambda$, the specific angular momentum. If we instead worked with the velocity potential $\Phi$, we would have written using the irrotationality condition that $hv_{\mu}=\nabla_{\mu} \Phi$. Since $v_{\phi}$ is nonvanishing, we could not have made $\Phi$ to be independent of $\phi$, unlike $\Psi'$, and accordingly, one obtains a three dimensional acoustic geometry.

If we allow for a $\phi$-dependence for $\Psi'$, we also expect to obtain a three dimensional geometry. We reserve this issue for a detailed study in future. In any case, studying only axisymmetric or $\phi$-independent modes for $\Psi'$ is nevertheless a reasonable assumption.    

Thus, it is clear from the equality of the tensorial parts that $f^{\mu\nu}$ bears the essential qualitative features of the causal structure of the internal acoustics, similar to the velocity potential approach. This is also manifest from Eq.~(\ref{metric3}). The non-relativistic limit of (\ref{metric3}) can be obtained by letting $r\to \infty$, $u_0,~c_{\rm s_0}\ll1$,
\begin{eqnarray}
f_{\mu \nu}\big\vert_{\rm NR} \equiv 
\frac{\left(1+\beta\right)}{u_0 c_{\rm s_0}^2 } \left[ \begin{array}{cc}
 u_0^2-\frac{c_{\rm s_0}^2}{1+\beta} & -u_0  \\ 
\\ -u_0 & 1 \\
 \end{array} \right].
\end{eqnarray}
Setting $\beta=0$ above recovers the result of~\cite{jkb-arnab-tapan-nabajit, unr81}.

To summarize, we have shown that the propagation of the $\phi$ independent linear perturbation of the mass accretion rate for axisymmetric flow can be described by a two dimensional matrix, the tensorial part of which bears similar acoustic causal structure as one gets via the perturbation of the velocity potential. We have shown this for the three accretion disk models in an equal footing. This generalizes the earlier results of~\cite{paper-I} for the spherical accretion with conical flow, and of~\cite{jkb-arnab-tapan-nabajit} derived in the non-relativistic scenario. 
Since the velocity potential and the mass accretion rate are apparently two very distinct quantities, both qualitatively and quantitatively, the demonstration of the association of the later with some acoustic causal structure seems far from obvious {\it a priori}. This is the main result of this work and to the best of our knowledge, this has not been shown earlier.

From a purely astrophysical perspective, it is important to know how $\Psi'$ behaves. From Eq.~(\ref{e35}) it is clear that
this behaviour is entirely determined by $f^{\mu\nu}$ (which is determined by the stationary state quantities) and the boundary conditions imposed upon $\Psi'$. 
Thus in a given scenario, if $\Psi'$ remains well behaved, we may conclude about the stability of the accretion process. 

To the best of our knowledge, the acoustic geometry by perturbing the velocity potential and related stability issues were first extensively studied in~\cite{M1980} in the Schwarzschild spacetime for spherical accretion. The stability of the accretion process was established via constructing bounded energy integrals and by studying the wave configurations generated by the perturbation of the velocity potential. In the following, as we have mentioned earlier, we shall instead use the acoustic geometry we have derived for the three disk models to determine the profile of the 
perturbation of the mass accretion rate. Precisely, we shall solve Eq.~(\ref{e35}) using (\ref{metric1}) corresponding to the possible standing and travelling wave configurations for $\Psi'$ to see if they are stable. We shall see that our results on stability are in qualitative agreement with~\cite{M1980} for the spherical/conical accretion disk model, whereas the vertical equilibrium model could indicate instabilities.

\section{Stability issues} 
\noindent
The general scheme to study stability of the accretion process is similar to that presented in e.g.~\cite{paper-I} and references therein, which discuss the stability issues for spherically symmetric flows. Here we proceed along the same line by applying this  scheme to axisymmetric flows for all the three disk models we have been concerned with, in order to check if we can predict any instability arising due to the nontrivial disk structure. We start by taking a $\phi$-independent (which we have assumed throughout) trial solution,
\begin{eqnarray}
\Psi'(r,t) = p_{\omega}(r) \exp(-i\omega t),
\label{s2}
\end{eqnarray}  
and substitute into Eq.~(\ref{e35}) to obtain
\begin{eqnarray}
\omega^{2} p_{\omega}(r) f^{tt} + i \omega \left[\partial_{r} \left( p_{\omega}(r)f^{rt} \right)+ f^{tr} \partial_{r} p_{\omega}(r)]-[\partial_{r} \left(f^{rr} \partial_{r} p_{\omega}(r) \right) \right] = 0.
\label{eqn4.52}
\end{eqnarray}
where $f^{tt}$ etc are given by Eq.~(\ref{metric1}).
We shall first consider the stability issues for standing waves, which 
requires $p_{\omega} (r)$ to vanish at two different radii, $r_1$ and $r_2$ at all times,
\begin{eqnarray}
p_{\omega}(r_{1}) =0= p_{\omega}(r_{2}).
\end{eqnarray}
Multiplying Eq.~(\ref{eqn4.52}) with $p_{\omega}(r)$ and integrating by parts between these boundaries ($r_1,r_2$), we get
\begin{eqnarray}
\omega^{2} = - \frac{\int f^{rr} \left( \partial_{r} p_{\omega}(r) \right)^2}{\int f^{tt} \left(p_{\omega}(r)\right)^2}.
\label{eqn4.29a}
\end{eqnarray}
We shall now use above the expression for $f^{\mu\nu}$ appearing in Eq.~(\ref{metric1}). We recall that in our notation (Eq.~(\ref{e1})) $g_{tt}$ is positive. Then since $c_{\rm s_0}^2<1$ always and we have $\beta\geq 0$, we have $f^{tt}<0$ always, and the denominator of Eq.~(\ref{eqn4.29a}) remains positive. So the nature of $\omega$ (i.e. whether real or imaginary) depends on the sign of $f^{rr}$. It is easy to show that $f^{rr}>0$ outside the critical point.

Usually the standing waves correspond to the subsonic flows~\cite{PSO1980} and we take the outer boundary $r=r_2$ at outside of the critical point.
Thus if the inner boundary $r=r_1$ is outside the critical point too, we have $f^{rr}>0$ and hence $\omega$ has two real roots confirming that the stationary solutions are stable. 

However, it is interesting to note from the expression of $f^{rr}$ that (Eq.~(\ref{metric1})), if the inner boundary lies inside the critical point (but outside the sonic point), $\omega^2$ can be negative (since $f^{rr}<0$ inside the critical point) and there can be either instability or damping effects, even though the flow is subsonic there (we recall once again that $\beta\geq0$).
Clearly, this happens only when $\beta\neq 0$. For any accretion disk model in which the weight function 
$H_{\theta}$ is not a function of the flow variables, we have $\beta=0$, and the critical and sonic points are coincident then (cf. discussions after Eq.~(\ref{e23})). In that case there will be no such damping or instability effects, because the inner boundary is located outside the sonic point.  This is in qualitative agreement with the result of~\cite{M1980}, derived for spherical accretion in the Schwarzschild spacetime, for the perturbation of the velocity potential.

Let us now come to the traveling waves. We use the trial power series solution similar to the flat spacetime~\cite{PSO1980}, 
\begin{eqnarray}
p_{\omega}(r) = \exp \left[ \sum_{n=-1}^\infty \frac{k_{n}(r)}{\omega^{n}} \right].
\label{eqn4.30}
\end{eqnarray}
In order to let the wave propagate to large radial distances, it is always necessary to have $\omega$ values to be large. In that case the advantage of making the above ansatz is that, it converges sufficiently rapidly. Thus the above ansatz is similar to the WKB solutions.

Substituting Eq.~(\ref{eqn4.30}) into Eq.~(\ref{eqn4.52}) and setting the coefficients of individual powers of $\omega$ to zero gives the leading coefficients in the power series, the first two of them being
\begin{eqnarray}
k_{-1} &=& i \int \frac{f^{tr} \pm \sqrt{(f^{tr})^{2}-f^{tt}f^{rr}}}{f^{rr}} dr, \nonumber\\
k_{0} &=& -\frac{1}{2}\ln(\sqrt{(f^{tr})^{2}-f^{tt}f^{rr}}).
\label{eqn4.32}
\end{eqnarray}
It can be explicitly checked from the expression of $f^{\mu\nu}$ that the corresponding $k_{-1}$ is purely imaginary. The leading
behaviour for the amplitude of $\Psi'$ is given by $k_0$,
\begin{eqnarray}
\vert \Psi' \vert \sim \left[ \frac{\Lambda \left(1+g_{rr}v_0^2\right)}{g_{rr}v_0^2c_{\rm s_0}^ 2\left(1-\frac{\lambda^2}{g_{\phi \phi}}g_{tt} \right)}\right]^ \frac14, 
\label{am}
\end{eqnarray}
The above amplitude remains bounded and ensures stability
of the perturbation if there is no turning point ($v_0= 0$). This requires $\lambda$ to be rather low (non-Keplerian). However, we note that for $v_0=0$, the mass accretion rate is not defined and our analysis is not valid in that case. For calculations for spherical accretion in the non-relativistic case, we refer our reader to~\cite{PSO1980}. Also, for analogous expressions for the perturbation of the velocity potential, see~\cite{M1980}.

All the above analysis were performed in the general background spacetime of Eq.~(\ref{e1}). Based on this, we shall now
briefly address below the explicit example of the Schwarzschild spacetime.
\subsection{The Schwarzschild spacetime}
\noindent
For the Schwarzschild spacetime the metric functions in Eq.~(\ref{e1}) are given by
\begin{eqnarray}
g_{tt} = g_{rr}^{-1} =  1-\frac{2M}{r},~ g_{\theta \theta} = \frac{g_{\phi \phi}}{\sin^2 \theta} = r^{2}.
\label{eqn4.23}
\end{eqnarray}
The explicit expression for the symmetric tensor $f^{\mu\nu}$ and its inverse $f_{\mu\nu}$ for the Schwarzschild background can be constructed from the general expression (\ref{metric3}). Recalling that we are working on $\theta=\frac{\pi}{2}$, we get 
\begin{eqnarray}
f_{\mu \nu} \equiv {\frac{\left(1-\frac{\lambda^2}{r^2}\left(1-\frac{2M}{r}\right)\right)^{-1/2}}{u_0 (1-u_0^2)}} \left[ 
\begin{array}{cc}
 \left(1-\frac{2M}{r}\right)\left(\frac{u_0^2(1+\beta)}{c_{s_0}^2}-1 \right)& \frac{u_0  \left( 1- \frac{1+\beta}{c_{\rm s_{0}}^{2}} \right)}{\left(1-\frac{\lambda^2}{r^2}\left(1-\frac{2M}{r}\right)\right)^{1/2}}     \\ 
 \frac{u_0 \left( 1- \frac{1+\beta}{c_{\rm s_{0}}^{2}} \right)}{\left(1-\frac{\lambda^2}{r^2}\left(1-\frac{2M}{r}\right)\right)^{1/2}}   &     \left(\frac{ \left(\frac{1+\beta}{c_{s_0}^2}-u_0^2 \right)-\frac{\lambda^2}{r^2} \left(1-\frac{2M}{r}\right)(1-u_0^2)}{  \left(1-\frac{2M}{r}\right)\left(1-\frac{\lambda^2}{r^2} \left(1-\frac{2M}{r}\right)\right)   } \right)
\end{array}\right].
\label{metric4}
\end{eqnarray}
The results for the standing wave analysis remain the same as discussed for the general case.

The amplitude of the traveling wave (Eq.~(\ref{am})) is now given explicitly by 
\begin{eqnarray}
\vert \Psi' \vert \sim \left[ \frac{\Lambda \left(1-\frac{2M}{r}+v_0^2\right)}{v_0^2c_{\rm s_0}^2\left(1-\frac{\lambda^2}{r^2}\left(1-\frac{2M}{r}\right) \right)}\right]^\frac14=   \left[ \frac{ \left(1+\beta\right)\left(1+ \frac{\left(1-\frac{c_{\rm s_0}^2} {1+\beta}\right)   \frac{\lambda^2\left(1-\frac{2M}{r}\right)}{r^2}} { \left(1-\frac{\lambda^2\left(1-\frac{2M}{r}\right)}{r^2} \right) \left(1-u_0^2\right)} \right)}  {c_{\rm s_0}^2 u_0^2 \left(1-\frac{\lambda^2\left(1-\frac{2M}{r}\right)}{r^2}\right) }\right]^\frac14,
\label{am2}
\end{eqnarray}
where in the last equality we have used Eq.s~(\ref{e20}) and (\ref{e32'}).
Eq.s~(\ref{am2}) is valid for the vertical equilibrium disk model, for which $\beta$ is given by Eq.~(\ref{e22''}).
In order to get the results for the other two models, we have to set $\beta=0$ in Eq.~(\ref{am2}) (cf. discussions at the end of Section 2).

Setting $\lambda=0$ and $\Lambda=1$ in Eq.~(\ref{am2}) recovers the result for the spherical flow, derived earlier in~\cite{paper-I, jkb-arnab-tapan-nabajit}. Expressions analogous to (\ref{am2}) was derived in~\cite{PSO1980} for non-relativistic spherical accretion process.

It is also interesting to note that the effect of $\lambda$ or the azimuthal flow is to magnify the amplitude of the wave. This can be understood as the centrifugal effect associated with the rotational energy of the flow.
Although it seems intuitively obvious, it is nevertheless essential to quantify such statement, as $c_{\rm s_0}$
may depend upon $\lambda$. Numerical analysis seems a suitable to address this issue.

\section{Summary and outlook}
\noindent
We have studied the linear perturbation and stability analysis of the mass accretion rate in general static, spherically symmetric spacetimes for flows with low angular momentum and discussed stability issues. We then applied the general formalism to the Schwarzschild spacetime. The motivation behind using the mass accretion rate surely lies in its direct physical relevance and its 
observational measurability for astrophysical accretion process~\cite{monika}. We have discussed different disk geometries for the accretion, which brings in subtle complexities and rich features not encountered for the radial flow~\cite{paper-I}.   
 
First, we proved that the perturbation effectively propagates through 
a symmetric 2-tensor which has acoustic causal properties qualitatively similar to that of one gets via the well known perturbation of the velocity potential~\cite{B1999}. This is one of the main findings of the present work and surely, such similarity is not obvious {\it a priori}, owing to the distinct properties, qualitative or quantitative, of these two quantities. 

Using this effective internal acoustic geometry, we next addressed the stability issues associated with the linear perturbation of the mass accretion rate. For conical or constant height model for the accretion disk, we have reestablished the qualitative features of stability of standing waves derived earlier in~\cite{M1980}, using velocity potential perturbation. For more nontrivial vertical equilibrium model, interestingly,
we have seen indication of damping or instabilities. For the traveling wave part, we have derived formal analytic expression for the WKB amplitude of the perturbation of the mass accretion rate, and qualitatively argued about its boundedness and hence stability. In other words, the chief difference of our stability analysis with that of existing literature, e.g.~\cite{M1980}, is twofold. Firstly, we have used the emergent acoustic geometry obtained via the linear perturbation of the mass accretion rate and not through the velocity potential. Secondly, we have studied the standing and travelling wave configurations for the perturbation of the mass accretion rate in the most general geometric configuration of the axisymmetric flow.

Our present work may lead to several interesting tasks. The first is to make this study for the Kerr background, which will surely bring in qualitatively new effects due to frame dragging, such as distinction between the pro- and retrograde orbits. Also, it will be highly interesting to make the qualitative or general existence statements made in the stability part (Section 4), especially regarding the standing wave and the amplitude of the travelling wave, quantitative. Perhaps this can be tried via numerical analysis. We hope to address these issues in our forthcoming works.

\vskip 0.4cm
\section*{Acknowledgement}
\noindent
Long term visit of DAB at Harish-Chandra Research Institute (HRI) has been supported by the planned project fund of the
Cosmology and the High Energy Astrophysics subproject of HRI. Majority of SB's work was done at HRI when he was a post doctoral fellow there. He acknowledges the ``ARISTEIA II" Action of the Operational Program ``Education and Lifelong Learning", co-funded by the European Social Fund (ESF) and Greek National Resources and IUCAA, India, for partial supports. He also sincerely thanks Jayanta K Bhattacharjee for useful discussions and encouragement.

\vskip.2cm
\section{The Bibliography}
\label{thebib}

\end{document}